# Evidence Based Decision Making in Blockchain Economic Systems:

From Theory to Practice

Marek Laskowski [†]

    Novera Capital, Toronto, Ontario, Canada, marek@novera.com

Michael Zargham

    Block.Science, Oakland, California, United States, zargham@block.science

Hjalmar Turesson

    Schulich School of Business, Toronto, Ontario, Canada, hturesson@gmail.com

Matt Barlin

    Block.Science, Oakland, California, United States, barlin@block.science

Danil Kabanov

    Insolar Technologies GmbH, Zug, Switzerland, danil.kabanov@insolar.io

Eden Dhaliwal

    Outlier Ventures, London, England, eden.dhaliwal@gmail.com

[†] Corresponding Author.

## ABSTRACT

We present a methodology for evidence based design of cryptoeconomic systems, and elucidate a real-world example of how this methodology was used in the design of a blockchain network. This work provides a rare insight into the application of Data Science and Stochastic Simulation and Modelling to Token Engineering. We demonstrate how the described process has the ability to uncover previously unexpected system level behaviors. Furthermore, it is observed that the process itself creates opportunities for the discovery of new knowledge and business understanding while developing the system from a high level specification to one precise enough to be executed as a computational model. Discovery of performance issues during design time can spare costly emergency interventions that would be necessary if issues instead became apparent in a production network. For this reason, network designers are increasingly adopting evidence-based design practices, such as the one described herein.

## CCS CONCEPTS

• Insert your first CCS term here • Insert your second CCS term here • Insert your third CCS term here

# 1  Introduction

Since the launch of Bitcoin in 2009, there has been increasing interest in blockchain and distributed ledger technologies. Valued by early adopters for its censorship resistance and decentralization (widely distributed peer-to-peer network participation), other properties of Blockchain networks are desirable in multi-stakeholder business relationships such as auditability, and byzantine fault tolerance [1]. The full design scope of these networks is an inherently interdisciplinary endeavor, incorporating perspectives from cryptography, distributed systems, computer science, game theory, and economics. A key challenge of decentralized system is designing incentive mechanisms that lead to reliable performance and self-sustaining activities within the network. A canonical and well documented example of this is Bitcoin. The key performance goal of Bitcoin is preventing the "double spend" problem in earlier digital cash efforts where a sender is able to simultaneously send a copies of the same digital money to two different recipients thusly creating cash out of thin air [2]. The Bitcoin network does a tremendous amount of work to ensure that Bitcoin is credited to be the rightful recipient and simultaneously nullifying the corresponding Bitcoin of the sender. A third party that is neither the sender nor recipient verifies this designation. That third party must be incentivized to verify accurately and not collude with the sender or the recipient. Roughly every ten minutes, the network selects a temporary leader ("miner") to add the next block to the Bitcoin blockchain, where the block is comprised of transactions broadcasted since the last block and verified by the peer-to-peer network to not be "double spent." Currently, a miner receives 12.5 bitcoins, or roughly $90,000 each time a new block is added plus a variable amount of transaction fees paid by senders. Both senders and recipients are assured that money cannot be transferred from insufficient funds and third-party miners are incentivized by the promise of receiving ("mining for") bitcoins to provide this assurance. The Ethereum network, the second most prominent public blockchain network, also has a similar token economics model. Ethereum is designed as a platform to execute distributed applications or so-called "smart contracts". These smart contracts as well as the data they operate upon are stored on the Ethereum blockchain. Like Bitcoin, miners validate new blocks and are rewarded with Ethers, the platform's cryptocurrency, and transaction fees paid by users for computation. Bitcoin and Ethereum are examples where economics is "baked into the protocol." That is, their respective cryptocurrencies are automatically generated and transferred by programmatic rules enforced by the peer-to-peer, network of stakeholders which are incentivized to collectively behave to sustain network operations.

By virtue of Bitcoin and other blockchain networks solving the double spend problem, an "Internet of Value" has emerged with thousands of blockchain networks today each with their own application-specific token models, many of which are ad-hoc inventions or tweaks of other models.

In general, this work is concerned with so-called "cryptoeconomics" [3] which describes the study of economic networks which form on top of blockchains or distributed ledgers that rely on cryptography. We use the term Token Economics to include Blockchains, DLTs, or decentralized networks built on different underlying technologies in which there are typically token-based economies, valued due to a limited or scarce supply of tokens which entitle holders to various privileges or utilities within the associated network.

The systematic development of token economies [4][5], sometimes referred to as token engineering, is an interdisciplinary endeavor incorporating perspectives from Systems Engineering, Complex Adaptive Systems (CAS), Economics & Game Theory, and Data Science. For instance, network dynamics models analytically prove that Bitcoin's specific token model leads the network to a stable, self-sustaining state [6]. The combination of system complexity and nascence of the field both make it unlikely that an initial ad-hoc attempt will discover an optimal configuration for a token economy model. For example, the Steem network has had 22 "hard-forks" or forceful network upgrades to since its launch in 2016 to adjust among other things the network token economics.

Therefore, the token engineering process is naturally iterative with phases of analysis, design, and deployment occurring in repeated cycles [5]. It's important to distinguish that this is the process for token modeling, not the wider software engineering process that leads to the development of the network software itself. The outputs of the token engineering process are parameterized models, results, and documentation that are implemented in the blockchain system.

The process includes several key phases:

- Token Model Analysis

At the outset, the system designers must answer which phenomena or Key Performance Indicators (KPIs) the blockchain network should maximize or minimize. During this phase, it's key to understand users and stakeholders of the blockchain application, specifically their motivations, behaviors, and inter-relationships.

- Token Model Design

The systems designer must answer how tokens are involved in maintaining the consensus and good governance of the network. In some networks, tokens are used for voting on consensus or network governance matters [7]. Often, a means of ensuring stakeholder accountability through mechanisms such as "staking" may be specified. Staking entails a participant posting a bond in order to take part in a network activity. The participant may forfeit the bond if they are judged to be a "bad actor" by other participants. Staking ensures that participants have "skin in the game" and are invested in network success [8]. Participant behavior that supports the objective function of the network can be incentivized through token rewards between participants and also through the minting of new tokens. This requires careful design of token emission or minting policies, which represent fiscal policies for the token system and underlying blockchain network. Finally, the application of Game Theory is essential in order to design mechanisms that drive participants toward Nash equilibria, or Schelling point solutions that seem natural or special to participants even without coordination amongst themselves.

- Token Model Deployment

Earlier on, blockchain networks launched with the implementation of the token economics model going live for the first time and managed by evaluating live network performance and making, often emergency, fixes on the fly. This led to a comparison of these networks to airplanes being constructed while in mid-flight. A maturation of the blockchain "ecosystem" has led to increasing use of simulation to bridge the micro characteristics (individual behaviour and transactions on the blockchain), to the emergent macro behaviour of the network as a whole (e.g. markets and coalitions). Ahead of putting a network into production with real valued tokens at stake, the simulation approach can mitigate risk and save resources and network value wasted through discovering flaws in production and deploying emergency fixes. In this context, a simulation model can be used to understand of how tokens are likely to move based on economic incentives, the so-called "token gravity" of the network [5].

\*\*\*

Given the potential impact of these decentralized networks over the coming years, we believe that Token Engineering is an area opportune for further investigation. In this paper, we present a real-life Data Science and Stochastic Dynamical systems modelling study conducted to develop a token economics model for Insolar, an 80-person blockchain startup with offices in five countries and headquartered in Switzerland. Insolar conducted an Initial Coin Offering (ICO) of their INS coins in December 2017 and raised $42 million [9]. As they recently ramped up their development and partnership efforts in preparation for their Main Network launch in 2020, Insolar worked with BlockScience (a design and consultancy firm focused on economic engineering) throughout 2019 to create a "digital twin" in order to validate and optimize their Token Economic Model.

In the present work, rather than present a comprehensive simulation study, we have a rare opportunity to provide a view of real-world application of token economics modeling to support a key design decision in a blockchain network prior to launch. To this end, our paper is structured as follows. In the next two section, we briefly describe the Data Science methodology employed, and the Stochastic Dynamical Systems modelling methodology respectively. Next, we further describe Insolar and detail their overall goals for the network which in turn directed the token economics modeling of an "application developer subsidy pool" - an incentive program to subsidize third parties who provide applications for users to execute on Insolar's MainNet public blockchain network. The application developer subsidy pool model is described in the context of the wider system model. Simulation results, insights gained, resulting design decisions, and validation of these decisions are presented to illustrate how the stochastic dynamical systems model is applied within a Data Science driven investigation. Finally, we provide concluding remarks about how the simulation informed Insolar's implementation of the token economics mechanism and provide general insights about cryptocurrencies and token economics.

## 2 Evidence-Based Decision Making and Design – a Data Science methodology

Although Data Science is a relatively new term, it is becoming synonymous with the application of advanced analytics, visualization, statistics, and related methods to provide knowledge and insights for organizational decision support [10]. Increasingly, simulation modelling is being used as a tool within a Data Science process especially when complex value networks are involved [11][12][13].

Simulation is complementary to other tools in the data science toolbox as a simulation model integrates assumptions and available data into an *in-silico* laboratory that encompasses the entire system being studied. Often, the modeling process itself is valuable in that it uncovers gaps in knowledge, contradictory assumptions, or under-specified parts of a system being designed [14]. For instance, to create the type of detailed model described in the following sections, designers must detail the system beyond the high-level business requirements found in whitepapers, down to fine-level control system logic that is necessarily consistent.

A famous process model for organizational evidence-based decision making is the Cross-Industry-Standard-Process or CRISP model [15]. Like the token engineering process described in the previous section, the CRISP model emphasises iteration, experimentation, and generation of new knowledge each time through the process loop. Indeed, the Token Engineering process described in the previous section can be mapped onto a simplified CRISP model as shown in Figure 1. In the figure, each step of the Token Engineering process is shown in italics alongside its corresponding phase of the CRISP model. Before the network goes live and is launched (inner loop), the initial assumptions of the system designers are incorporated to create a Token Economic Model or design which is in deployed within a simulation or "digital twin" of the network, and its performance evaluated. The results, data, and knowledge produced by this simulation are then used to inform the Business Understanding or Analysis for the next iteration through the Token Engineering process. The modeling process itself, along with the analysis of the results, identifies issues which can be dealt with at design time; issues which would otherwise be costly to deal with in production. Once designers have gone a few times through the cycle, the design has been validated and optimised enough to launch as a live system. The Post-launch (outer loop), the simulation model can be parameterized using data observed on the live network, the model used to better understand the health of the network, and to conduct "what-if" analyses [14] in order to evaluate proposed refinements to the design of the network.

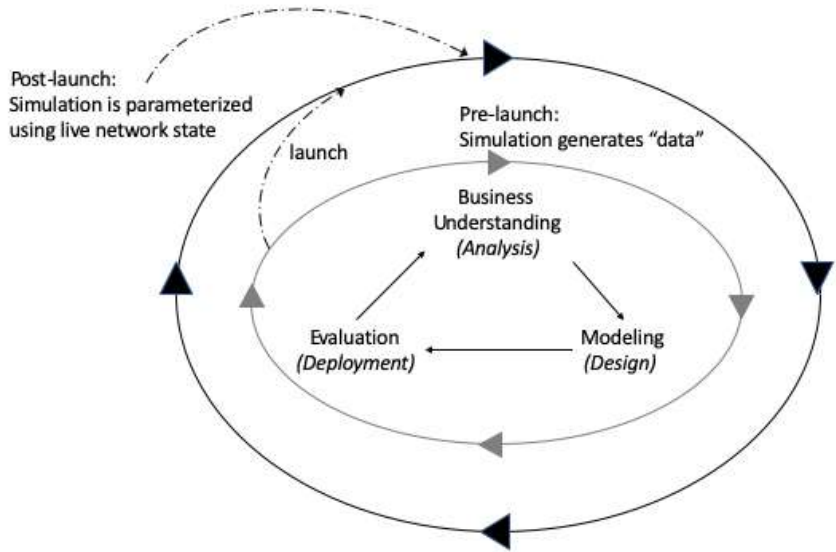

**Figure 1. The Token Engineering process mapped onto a simplified CRISP data science process model. Token Engineering phases are shown in *italics* alongside the corresponding phase of the CRISP model.**

Bitcoin was designed to be deliberately simple in order for its functions and mechanisms to be reasoned about without requiring deep analysis. Increasingly blockchain networks are launching with more complex use cases than "digital gold" or "peer-to-peer cash" and therefore involve increasingly complex mechanisms and performance indicators. Not surprisingly, network designers are increasingly turning to evidence from simulation modeling in order to support their decision making with respect to network design.

The required modeling framework must be capable of:

- Simulating temporal dynamics
- Multi-scale modeling of network, ledger transaction, and economic levels
- Integrates micro-foundational and meso-institutional characteristics into the macro-observable system properties.

Such a framework is described in the next section.

## 3 The Statespace Framework: An Event Based State Space Model

In this section we describe a previously developed dynamic system modeling approach useful for decision making in a cryptoeconomics context [16]. Here we introduce a minimal theory relevant to the understanding of the present study. For a more complete treatment, please see [16] and [6].

The approach has its theoretical roots in Differential Games [17] which itself spans the disciplines of control theory and game theory. Our approach is further inspired by Ole Peters and Alexander Adamou's work on ergodicity economics [18] and John Sterman's work on business dynamics [19].

Unlike traditional control theory problems, when designing an economic network of human agents and incentives, designers at best have indirect control over the incentive structure with little control over the exact behavior [6] of system components and agents.

A stochastic dynamical system model enables us to represent complex relationships within a system including the interaction of agents that follow their own policies as a function of their own beliefs of the system state and their own utility function. The collective behavior agents result in secondary and tertiary system dynamics which are often nonlinear and hold surprises for system designers. The role of the designer, therefore, is to arrive at a set of rules and incentives so that the system level goal, as determined earlier, will be robust to the unpredictable individual actions of individual agents.

Figure 2 presents a dynamic systems model for agent's beliefs and policies [6]. External, stochastic processes are fed into agent polices, collectively update the system state, which is only partially observable from the perspective of the agents. The agents within the system can then observe the state and form of their beliefs, signals, and objectives that are the basis for feedback loops and polices at the next iteration. With regards to agent behavior, private signals are denoted $\sigma$, private goals are denoted $V(\cdot)$, as well as a partially observable state $\mathbf{X}$ with an observable subspace $\mathbf{Y} \subseteq \mathbf{X}$ and some additional stochastic processes $\delta$ representing input from the environment drawn from potentially unknown distributions

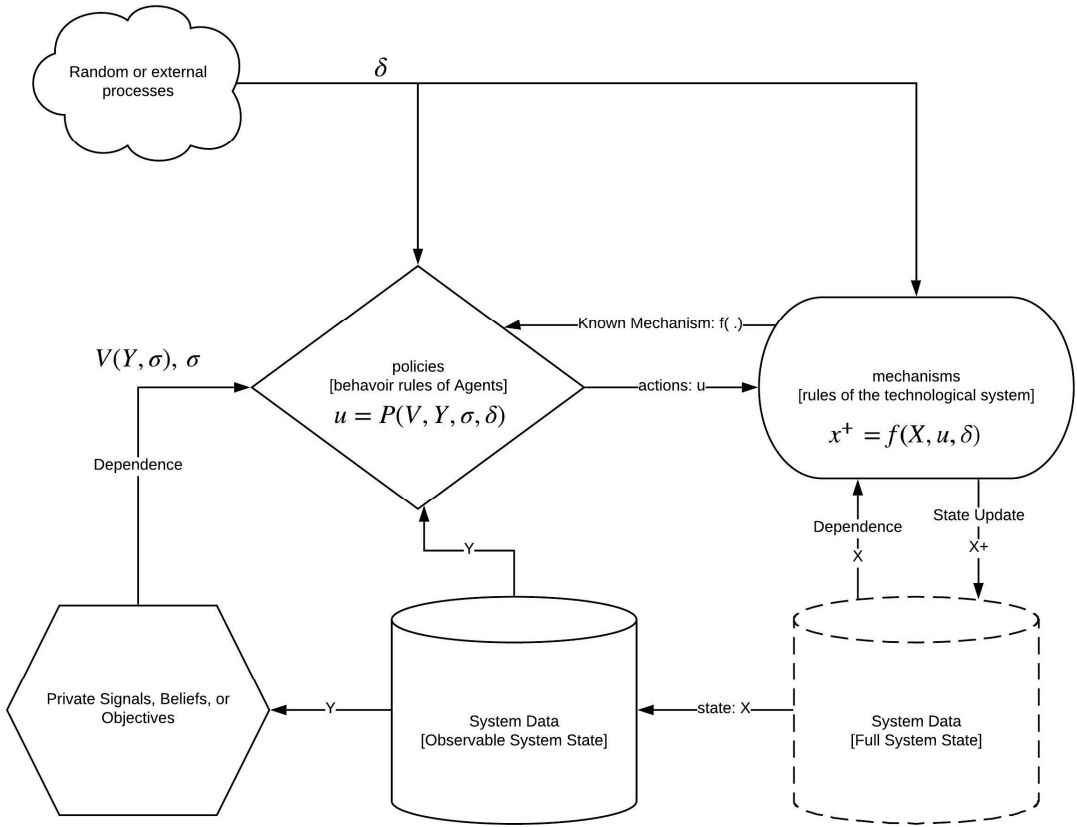

**Figure 2. An expanded representation of the differential game which accounts for states, actions, and the feedback loops inherent in the system [16].**

The framework also borrows from a modeling framework used in control engineering known as "state space representation" [20]. Within this formulation there is a set of state variables evolving over time according to a set of rules. If we assume that time is modeled discretely, the rules by which the system state evolves can be described as a first-order difference equation, in which the values of the states at a given time $t \in \mathbb{N}$ are completely dependent on the values of the state variables at $t-1$. If the initial state of the system is known at $t=0$ then the state at all times $t>0$ can be determined by solving the aforementioned recursion. We propose a linear state-space model of a blockchain network whose set of states represent transaction addresses. We propose a linear time-expanding (LTE) state-space model which we use the analyze the behavior of the blockchain network as it evolves through time.

With reference to Figure 2 the core of our computer aided design tools contains discrete time differential equations in a state space, however state variables are real-valued. The following definitions are used to characterize the unbounded agent dimensional state space and to relate that bounded state space over which system requirements may be defined. This model does not require that the system be implemented with a centralized state, rather the mathematical characterizations of the macro scale evolution of the system are in terms of formally defined micro elements. This framework is critical for understanding the relationships between agents and accounts within a decentralized system. Furthermore, the timescale of the state space model is defined such that agent actions action can be properly modeled. A discretization of the sequence and aggregation of actions into transaction blocks is necessary to simulate the system at this level. However, the construct provided is sufficiently general that if the discrete time is mapped to atomic events, the differential equations models may be used for event-based simulation without deriving new models.

This model is defined over addresses. All addresses are defined by public key private key pairs, where private keys are used for cryptographic proof of the right to act as agent. Without loss of generality more complex schemes such as multi-signature schemes can be substituted for the simple private key proofs assumed here. Further characterization of cryptographic schemes is not required as this section focuses on the system dynamics rather than the means of enforcing those dynamics.

>**Definition 1** Let $\mathcal{A}$ be the set of all possible **Addresses** as determined by the range of the cryptographic hash function used in the system implementation. At any time the set of addresses that exist is $\mathbf{A} \subset \mathcal{A}$

An address $a \in \mathbf{A}$ is called an agent when referring to the address which takes an action. An address $a \in \mathbf{A}$ is called an account when it contains code declaring one or more states and associated mechanisms which are exposed to other agents. All accounts are agents, but not all agents are accounts.

>**Definition 2** The **Ledger State** is a shared data structure $\mathbf{L} \in \mathcal{L}$ which evolves when agents perform transactions by taking actions with respect to mechanisms; $\mathcal{L}$ denotes the space of all valid ledger states $\mathbf{L}$. The state $\mathbf{L} = \{\mathbf{X}, \mathbf{T}\}$ where $\mathbf{T}$ us a (partially) ordered list of transactions. Each $t \in \mathbf{T}$ is defined by an agent $a \in \mathbf{A}$, a mechanism $f \in \mathbf{F}$ and an action $u \in \mathcal{U}$.

The list of transactions $\mathbf{T}$ need not be strictly ordered because transactions are local operations on the state with respect to accounts. If it suffices to know the order of any transactions with dependence on or modification of a shared element of the state $x \in \mathbf{X}$. Further formalization of event orderings is deferred from discussion in this document.

>**Definition 3** Consider the **Local State** to be a subset of $X_i$ which is a subset of the global state $\mathbf{X}$ that is declared by account $i \in \mathbf{A}$. The address $i$ is both agent and account so it has local state $X_{ii}$ but other agents may control elements of $X_i$. The elements of the state declared by $i \in \mathbf{A}$ but controlled by $j \in \mathbf{A}$ are $X_{ij}$.

It is immediate that each local state $X_i$ must be declared by a unique account $i \in \mathbf{A}$. Therefore, the local states can be interpreted as a partition of $\mathbf{X}$ over accounts. It is not, however, assumed that there is a partition over agents dimension.

>**Definition 4** Consider the set of **Mechanisms** to be $\mathcal{F}$ such that any $f \in \mathcal{F}$ is an operator

$$f: \mathcal{X} \times \mathcal{U} \longrightarrow \mathcal{X} \tag{1}$$

Where $\mathcal{X}$ is the space of all possible states **X** and $\mathcal{U}$ is a space of all legal actions associated with the particular mechanism $f$.

Mechanisms like states, must be declared by an account and in many cases will have been declared alongside specific local state variables $X_i$ which the mechanism operates on. However, no such assumption will be made formally.

**Definition 5** *For any mechanism $f \in \mathcal{F}$, for any agent $a \in \mathbf{A}$ there is a state dependent **action space** representing all legal actions for agent $a$ given a state **X** under mechanism $f$. A particular **action** is denoted $u \in \mathcal{U}$.*

**Definition 6** *The set of all possible transactions is denoted $\mathcal{T} = \mathcal{A} \times \mathcal{F} \times \mathcal{U}$ where an element $t \in \mathcal{T}$ is defined $t = (a, f, u)$. In order for the transaction to be valid, agent $a$ must have the right to perform the state update operation $\mathbf{X}^+ = f(\mathbf{X}, u)$ given the current state **X**.*

As previously defined in the Ledger state, a sequence of transactions organized into a (partially) ordered list is denoted **T**. The partial ordering may be considered without loss of generality by noting that such a partial ordering is defined precisely by the independence of the final output to the ordering of the list. This occurs when there is strong separation in the states, accounts and mechanisms involved in the transactions in **T**.

**Definition 7** *A **policy** $P: \mathcal{X} \longrightarrow \mathcal{U}$ is a state dependent strategy over a particular mechanism $f \in \mathcal{F}$. An agent $a \in \mathbf{A}$ is said to be using policy $P$ over mechanism $f \in \mathcal{F}$ if it monitors the state $\mathcal{X}$ and broadcasts transaction $t = (a, f, u)$ associated with action $u = P(x) \in \mathcal{U}$.*

This definition allows for the case that no transactions are made because the conditions for a transaction under policy $P$ are never met but it excludes the degenerate cases where there is no state for which a transaction will be generated. A practical consideration of this framing is the agents engagement level in monitoring the state. For the purpose of simple notation the sampling rate of the monitoring process is absorbed into the definition of $P$. During simulation. it is made explicit for tuning purposes.

**Definition 8** *Consider ledger **State Transitions**; the Ledger State may be updated for any valid sequence of transactions $\mathbf{T} = [..., t, ...]$, where $t = (a, f, u)$ is valid given **X** when the operation is applied. Without loss of generality, it is assumed all transactions are valid because invalid transactions are rejected.*

An atomic update is defined
$$\mathbf{X}^+ = f_t(\mathbf{X}, u_t) \text{ for any valid } (a_t, f_t, u_t) \text{ given } \mathbf{X} \tag{2}$$

where $f_t$ is the mechanism used in transaction $t$ and $u_t$ is the action taken for transaction $t$. For a block defined by sequence of transactions **T**., the state update is
$$\mathbf{X}^+ = f_N(f_{N-1}(\cdots f_1(f_0(\mathbf{X}, u_0), u_1) \cdots, u_{N-1}), u_N) \tag{3}$$

where the list of transactions **T** is indexed by $\{0, 1, ..., N\}$.

**Definition 9** *The **Ledger Trajectory** is a sequence of ledger state $\mathbf{L}(k) = \{\mathbf{X}(k), \mathbf{T}(k)\}$, indexed by $k = \{0, 1, ..., K\}$ such that*

$$\mathbf{X}(k+1) = f_N(f_{N-1}(\cdots f_1(f_0(\mathbf{X}(k), u_0), u_1) \cdots, u_{N-1}), u_N) \tag{4}$$

*for transactions $\mathbf{T}(k)$ is indexed by $\{0, 1, ..., N\}$.*

Notation may be simplified by defining $F_k$ to be the composition of the transactions in $\mathbf{T}(k)$ such that
$$\mathbf{X}(k+1) = F_k(\mathbf{X}(k)) \tag{5}$$

denoting the closed loop state update accounting implicitly for the actions $u = (\mathbf{P}(X))$.

At any time the most recent Ledger State may be denoted $\mathbf{L}(K) = \{\mathbf{X}(K), \mathbf{T}(K)\}$ where the integer K is the block height and $\mathbf{L}(0) = \{\mathbf{X}(0), \mathbf{T}(0)\}$ is the genesis block. The number of transactions is dependent on the block $N = \mathrm{N}(t)$. Under this definition the Blockchain is characterized precisely by the trajectory of generalized dynamical system in canonical form. As defined, the differential equations can be used to characterize the system with atomic transactions as the basic unit of time. Organizing transactions into blocks provides a means of testing block-based logic.

Any mechanism that can be implemented as an account under this framework provides an explicit contribution to the actions available to all other accounts within the system. The explicit characterization of an account and its subsequent state changes permits the estimation of changes in any utilities defined over the network state. Using the second order discrete networked system model, it is possible to both formally analyze the reachable state space and simulate the response to incentives with respect to a variety of behavioral assumptions.

The implementation of this framework is a simulation platform named cadCAD (Computer-Aided Design for Complex Adaptive Systems)[1] Its features include tight integration with the python data science stack (e.g. scikit-learn, numpy, scipy, matplotlib), and its amenability to Monte Carlo simulations.

## 4 A Use case: Insolar Blockchain

We present a case study conducted to develop a token economics model for Insolar, an 80-person blockchain startup with offices in five countries and headquartered in Switzerland. Insolar conducted an Initial Coin Offering (ICO) of their INS coins in December 2017 and raised $42 million [9].

During the latter stages of development, Insolar has been testing its blockchain software in the context of small, permissioned network pilots. The in-development version of Insolar's platform is being used, for example, to operate a neighborhood micro-grid in Toronto, Canada in addition to other projects in renewable energy, supply chain management, and mining and natural resources.

Insolar's strategy is to become an enterprise-focused, permission-less blockchain that will provide a trust-layer between individuals and small & large enterprises. The specific stakeholders in the MainNet token economy include: Individual and enterprise users, who will use and pay for applications (smart contracts) that are deployed on the MainNet; Application Developers, who create and deploy applications (smart contracts) and receive fees for their usage; and Resource Providers who make available hardware capacities for running applications and receive fees for this provision.

Once MainNet launches, the currently circulating INS tokens (ERC-20 tokens, on Ethereum Main Net) will be converted 1:1 to XNS tokens on the Insolar Main Net. The consensus algorithm on Insolar MainNet is Proof-of-stake (PoS) and will use the XNS token to stake for securing the network. The XNS tokens that are staked serve as a source of insurance for the MainNet: If users, Application Developers, or Resource Providers violate Service Level Agreements (SLA's) with each other, coins from the pool of staked funds can be used to pay the rightfully aggrieved party. Stakers, in turn, are rewarded a staking fee – an investment income, akin to interest, earned from leaving their staked funds backing or otherwise facilitating economic activity.

In general, blockchain system designers have multiple objectives to simultaneously optimize for: firstly at the ledger layer to keep costs low for participants, while simultaneously maintaining incentives for network operators over time. At the marketplace economic layer that emerges above the ledger layer designers want to maximize efficiency and participation ultimately increasing the market size. Insolar decided to incorporate an evidence-based decision approach in designing the mechanisms that underpin the economics of the MainNet so that they meet the aforementioned overall objectives. More precisely, applied systems dynamics modelling and simulation to finely tune MainNet's token economics model.

---

[1] https://github.com/BlockScience/cadCAD

Beginning with whitepapers and existing documentation, the precise behaviour of the system mechanisms and other features of the Token Model were specified precisely enough to implement in the computational environment mentioned earlier (cadCAD). This back-and-froth process involving all stakeholders took place over several weeks spanning April – May 2019 to complete. The specified model was concurrently implemented on top of the cadCAD framework as decisions were made, with major implementation efforts completing in early June 2019. The rest of June was spent verifying and validating the simulation model, a step that would be necessary before using the simulation model to provide evidence for design decision support. A full simulation study including model validation and verification is beyond the scope of this paper, and will be published elsewhere, however for illustrative purposes, some results which add credibility to the model are presented in the following section.

Generally, in this phase of the blockchain network's lifecycle validation and verification is performed through repeated trials of parameterization, execution of the model, observing the macro outcomes, and analyzing whether the outcomes match expected or well understood, or reasoned phenomena. The latter is judged by expert stakeholders who are ultimately responsible for determining whether the simulation model is an acceptable representation or digital twin of the network undergoing design by reasoning about model output given a set of initial parameters.

Then, in late June and onwards, the validated model was used to analyze the effect of all mechanisms within the Insolar blockchain economy. Exhaustive simulations were carried out to understand the effects of applying mechanisms to varying degrees, spanning hundreds of pages of graphs and analysis. Through this exploratory process, a key decision point arose; how to best incentivize pioneering Application Developers to contribute applications while the MainNet is nascent and network effect has not been achieved yet. An Application Developer Subsidy was proposed to incentivize developers. Initially, the goal of the investigation was to determine how much should be put into the Application Developer Subsidy Pool. Overallocation to the pool could represent an inefficiency as the tokens could be better put to work elsewhere; and under-allocation could stifle network growth. It was decided that a key performance indicator (KPI) in this case would be the XNS price as it reflects a health and growing economy, while other indicators are looked at to ensure validity of the experiments.

## 4.1 Model details

Figure 3 illustrates the token flows corresponding to costs incurred by an application developer as well as application subsidies. Application Developers have to pay fees to resource providers for CPU cycles, bandwidth, and database provisions required to run their applications. They also pay the Insolar Foundation Resource Platform Fees to deploy applications the MainNet. Whereas these fees are paid for by an application provider grossly, there are application platform fees charged on a metered basis per deployed application; the more a particular application is executed, the more fees have to be paid to the MainNet. The Resource Platform Fees can be thought of as a type of "tax" on Resource Fees paid to Resource Providers. In order to help foster the organic growth of the Insolar network ecosystem, the Insolar foundation will offer the Application Developer Subsidy paid in XNS token incentives to build and deploy applications on the Insolar platform. Once received by the Application Developer, subsidies can be spent however the Application Developer sees fit, for example to pay Resource Fees to Resource Providers. The application developer subsidy is initially funded with a seed amount of XNS tokens which are released from a treasury at an exponential decay rate. However, this subsidy pool is replenished with the application platform fees that are collected.

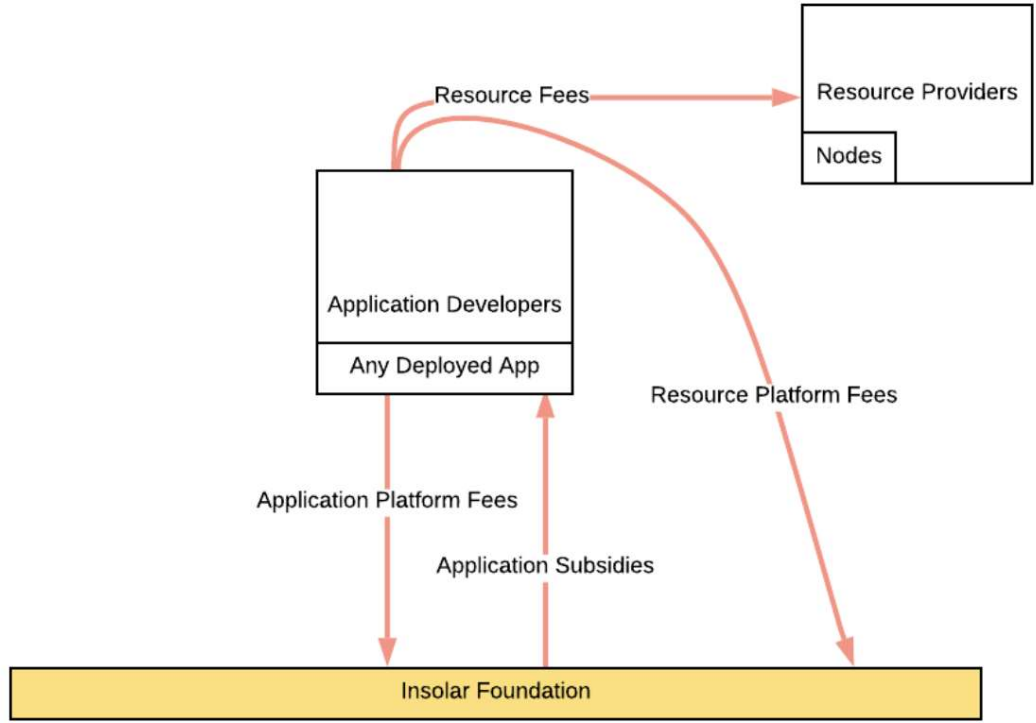

**Figure 3. Fees and subsidies related to Application Developers within the Insolar public MainNet ecosystem**

The desired effect of this subsidy may include: 1) Increased number of applications deployed: 2) Increased resources used; 3) Increased users; 4) Increased platform fees; 5) Increased demand for XNS; and 6) Increased price of XNS. The increased price of XNS is reflective of the market demand for a fixed supply of tokens available within the economy.

The number of tokens distributed to Application Developers at a particular time, *A(t)*, through the Application Developer Subsidy is governed by the initial reward pool size $A_0$, and the exponential decay rate, λ. The differential equation governing the change in subsidy over time is therefore:

$$\frac{dA}{dt} = -\lambda \cdot A \tag{6}$$

Through integration this results in.

$$A(t) = A_0 \cdot e^{-\lambda t} \tag{7}$$

In addition to these fees and subsidies, application users may pay application developers additional usage fees, however these are optional and not shown in the figure as they are not considered as part of the analysis here.

## 5  First Iteration of Application Developer Subsidy Simulations

In this section we present an illustrative "slice" of the token economic design process, to demonstrate how simulation models fit into the Token Engineering process, rather than presenting a comprehensive simulation study. As mentioned earlier, the initial goal of the designer was to understand how many XNS tokens to allocate to an Application Developer Subsidy Pool. The efficiency of this choice is assessed indirectly through changes to the XNS token price, as an overall indicator of network health, efficiency, and value.

A simulation study was performed to better understand the effects of varying the initial amount tokens devoted to the developer subsidy. The initial number will consequently affect the absolute rate of XNS token dispersal. The simulation scenarios considered in this initial investigation are summarized in Table 1. As mentioned, additional experiments were run during the design and analysis phase, which include but are not limited to exploring the model behaviour for validation and verification as well as evaluating the effects of other key Token Model parameters. These other experiments are omitted here for brevity.

The simulations were carried out in a Monte Carlo fashion using four possible scenarios for the starting reward pool size, and the mean of the resulting observed variables were computed. Although explored later, the Decay Rate remains constant throughout these experiments, as do the number of simulated time steps, and Monte Carlo runs. The key output state variables in the model include subsidies paid to application developers as well the predicted price of XNS in dollars.

**Table 1: Application Developer Subsidy Simulation Scenario Parameters – First**

| Initial Reward Pool Size (XNS) | Decay Rate | Time Steps (days) | Monte Carlo Runs |
| --- | --- | --- | --- |
| 250 x 10e6 | 0.0005 | 3652 | 100 |
| 500 x 10e6 | 0.0005 | 3652 | 100 |
| 750 x 10e6 | 0.0005 | 3652 | 100 |
| 1000 x 10e6 | 0.0005 | 3652 | 100 |

The outcomes that we plotted independently include:
1. The Application Developer Subsidy pool at any given time. Counted in XNS tokens (Figure 4)
2. The accumulated XNS tokens flowing to application developers (Figure 5)
3. The converted (dollars) income of XNS tokens to application developers (Figure 6)
4. The predicted XNS price in dollars (Figure 7)
5. The number of XNS tokens held by the foundation treasury (Figure 8)
6. The value of XNS tokens held by the foundation treasury (Figure 9)

In this initial investigation we began by gathering and considering evidence that the model is functioning correctly.

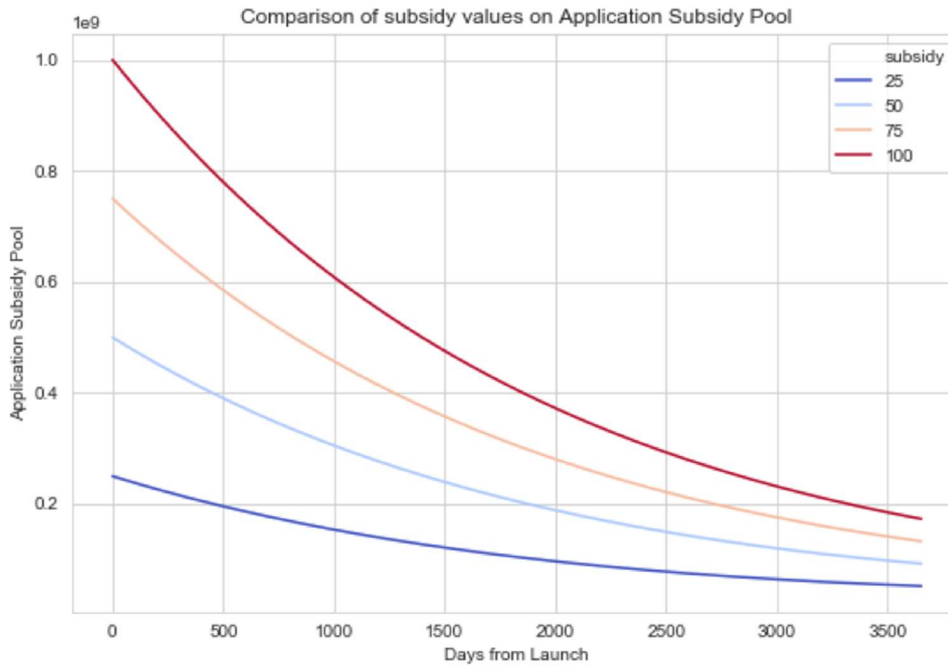

**Figure 4. The Application Developer Subsidy pool size at any given time (XNS Tokens)**

For instance the results presented in Figure 4 provide evidence that the model is functioning correctly. The expected exponential depletion of the application developer subsidy pool over time is exhibited by the trends in Figure 4 which confirm the initial starting reward pool size followed by exponential decay in token distribution over time. It is clear from the figure that the decay rate has been chosen such that there is a "long tail" for the distribution strategy in that it continues to distribute significant numbers of subsidy tokens for at least a decade following launch.

The resulting XNS token distribution to the application developers can be seen in Figure 5, which is again intuitively reasonable given the described exponential decay model and varying initial reward pool size. This figure confirms the correct operation of the model, in distributing Subsidy Pool Tokens to Application Develoeprs.

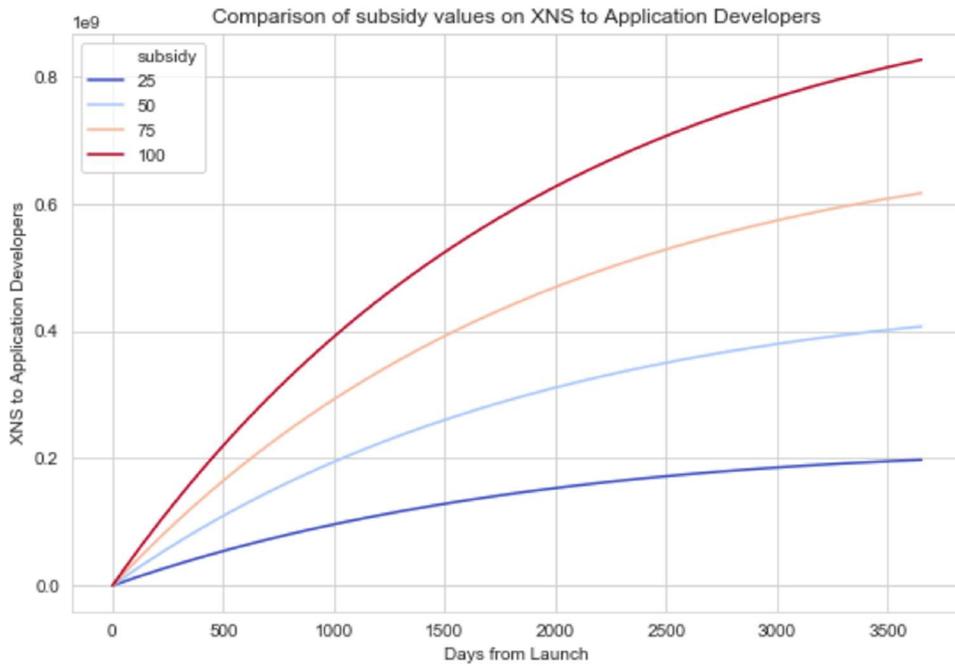

**Figure 5. The accumulated XNS tokens flowing to application developers**

Interestingly, the net effect on the cumulative income to application developers converted to dollars shown in Figure 6 is less intuitive or obvious. Although, the greatest subsidy does yield the most benefit to application developers, converse is not necessarily true after 10 years of projected network history. Clearly, there exists a complex and non-linear relationship between the initial subsidy pool size and reward value delivered to application developers.

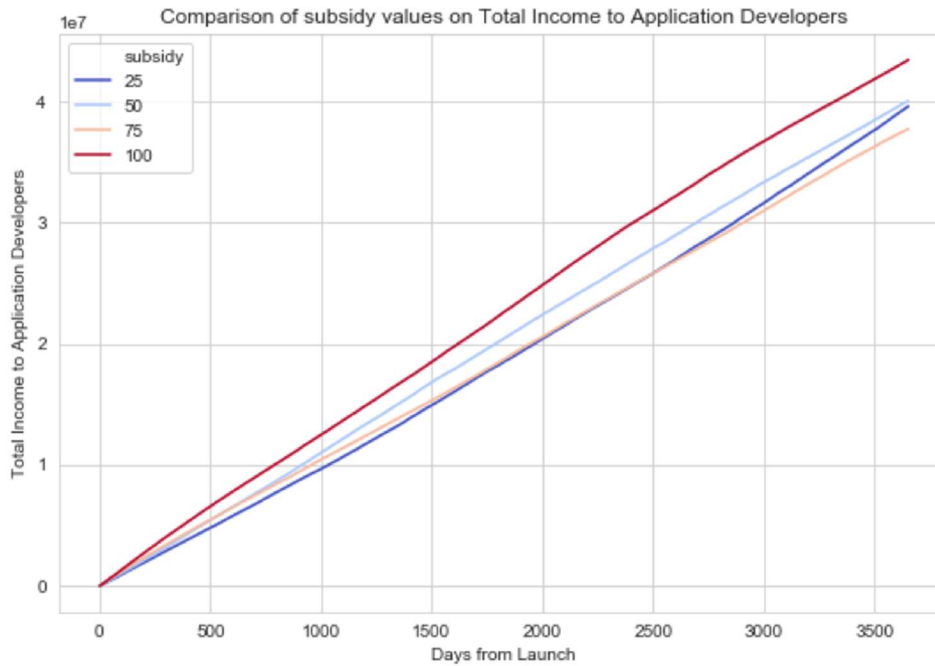

**Figure 6. The converted (dollars) income of XNS tokens to application developers**

To gain some further insight into why this relationship exists, we can look at projected XNS token price for each initial reward pool size scenario. As shown in Figure 7, the larger the initial reward pool size, the higher the generally resulting token price. However, we note a couple of things of interest. One is that diminishing returns are observed; doubling the tokens available in the subsidy pool does not have an equal effect on the price of XNS. Another interesting thing to note, is that the lowest subsidy amount appears to have a negative effect on the price. But most importantly, there are some interesting temporal dynamics observed within the first three years; some strategies deliver better results early on, but are eclipsed in the long run.

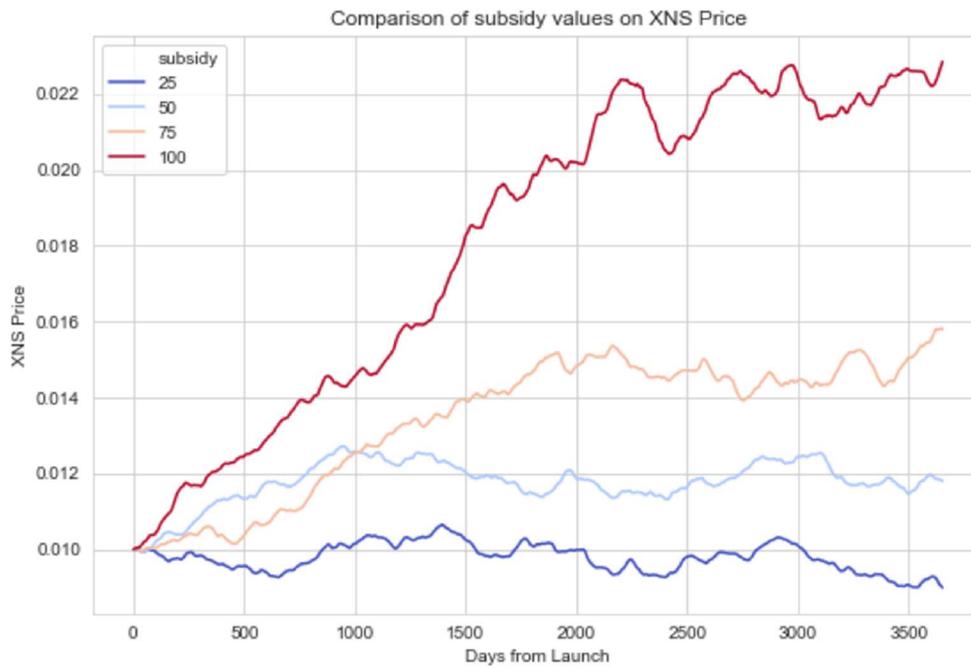

**Figure 7. The predicted XNS price in dollars**

At this stage of investigation, the designers also considered the initial reward pool size effect on the value of the tokens held by the foundation. Predictably, the foundation's holdings of XNS tokens vary inversely with initial application subsidy pool size as shown in Figure 8.

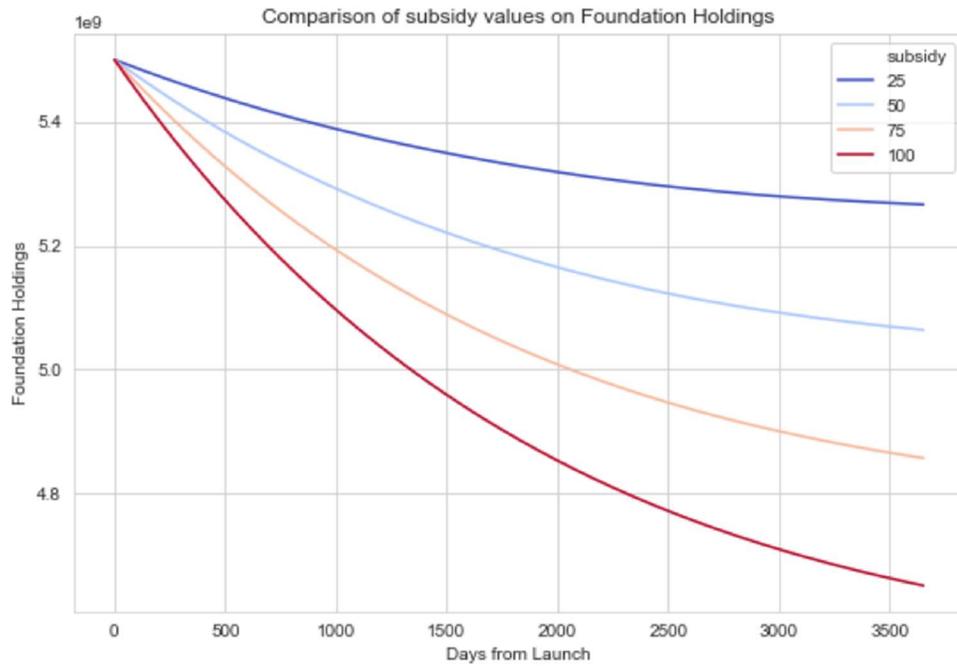

**Figure 8. The number of XNS tokens held by the foundation treasury**

However, due to the XNS price action demonstrated earlier, the value of tokens held by the foundation does increase as initial subsidy pool size is increased as evidenced by Figure 9. Once again, we observe an interesting temporal relationship suggesting something interesting is going on in the first few years of projected network history.

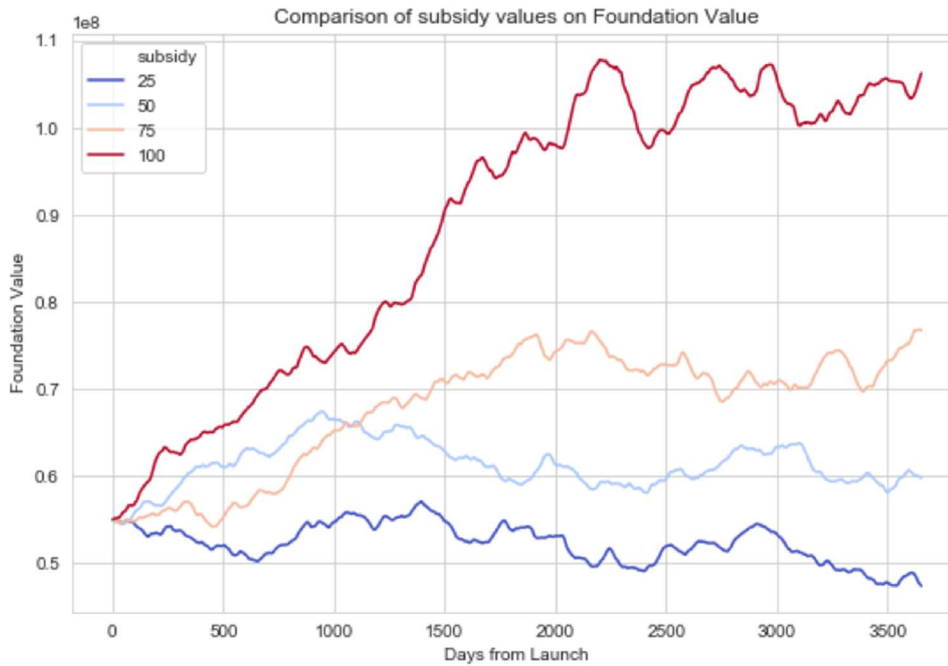

**Figure 9. The value of XNS tokens held by the foundation treasury**

## 6   Discussion – First Iteration

As suggested by the phenomena presented in figure 7 and confirmed by the phenomena observed in figure 9 a complex nonlinearity exists in the first 2-3 years of simulated network history. This drew designers to the realization that as a complex system, relatively small changes in the early history of the network could have a disproportionate effect later. As a result of this observation, and the ensuing design discussions, it was decided that instead of distributing the application developer subsidy in a "long tail" fashion over 10 years, the subsidy would be paid out earlier in the networks lifetime when it has more chance to make a positive impact, targeting the first two years of network operation. The Token Model design was therefore refined using the new distribution objectives. The "digital twin" simulation model was updated using the new Token Model, and used to test the effects of this new approach. This second round of investigation is described in the next section.

## 7 Second Iteration of Application Developer Subsidy Simulations

Verification and validation of the updated "digital twin" simulation model was again performed by expert stakeholder consensus, this time also considering the expected change in behaviour from the previous Token Model to the updated Token Model with a shorter distribution schedule. Changing the distribution strategy to focus on the early network history period was accomplished by changing the Decay Rate, λ, to 0.01. A secondary goal was to understand whether changing distribution strategies would make even smaller initial reward pool allocations viable. Therefore, two lower values for initial reward pool size were investigated, as summarized in Table 2, below.

**Table 2: Application Developer Subsidy Simulation Scenario Parameters – Second**

| Initial Reward Pool Size (XNS) | Decay Rate (λ) | Time Steps (days) | Monte Carlo Runs |
|---|---|---|---|
| 10 x 10e6 | 0.01 | 3652 | 100 |
| 50 x 10e6 | 0.01 | 3652 | 100 |
| 250 x 10e6 | 0.01 | 3652 | 100 |
| 500 x 10e6 | 0.01 | 3652 | 100 |
| 750 x 10e6 | 0.01 | 3652 | 100 |
| 1000 x 10e6 | 0.01 | 3652 | 100 |

The outcomes that we plotted independently include:
1. The Application Developer Subsidy pool at any given time. Counted in XNS tokens (Figure 10)
2. The accumulated XNS tokens flowing to application developers (Figure 11)
3. The predicted XNS price in dollars (Figure 12)

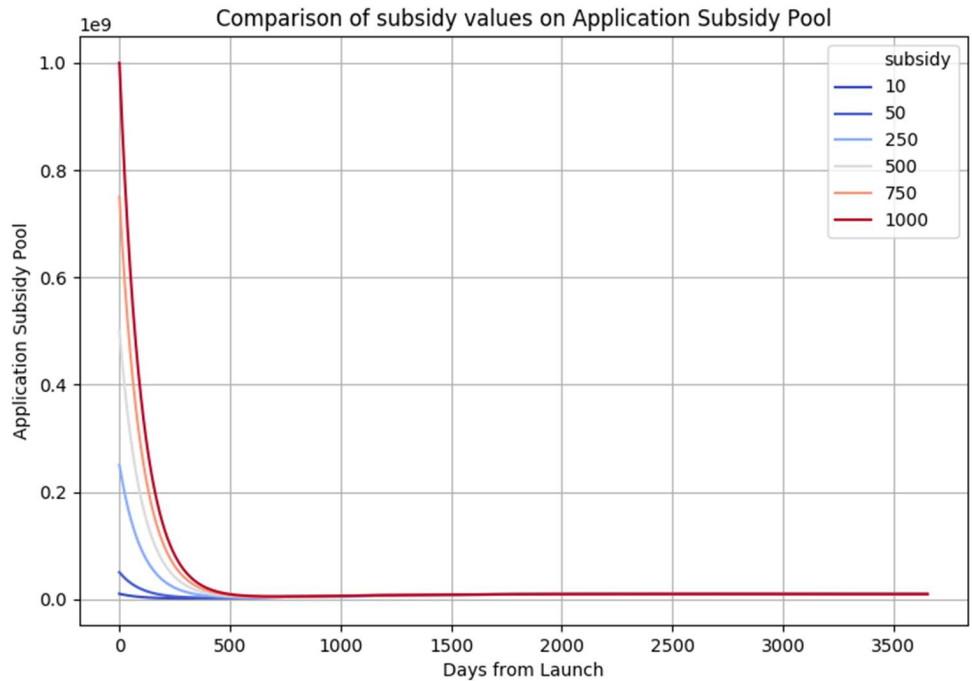

**Figure 10. The Application Developer Subsidy pool size at any given time (XNS Tokens) – revised developer subsidy distribution strategy.**

Figure 10 confirms that indeed, the vast majority of application developer subsidies are distributed within the first two years. Figure 11 verifies that this value is flowing into developer wallets. These two figures are plausible when compared to the behavior of the system as presented in figures 4 & 5.

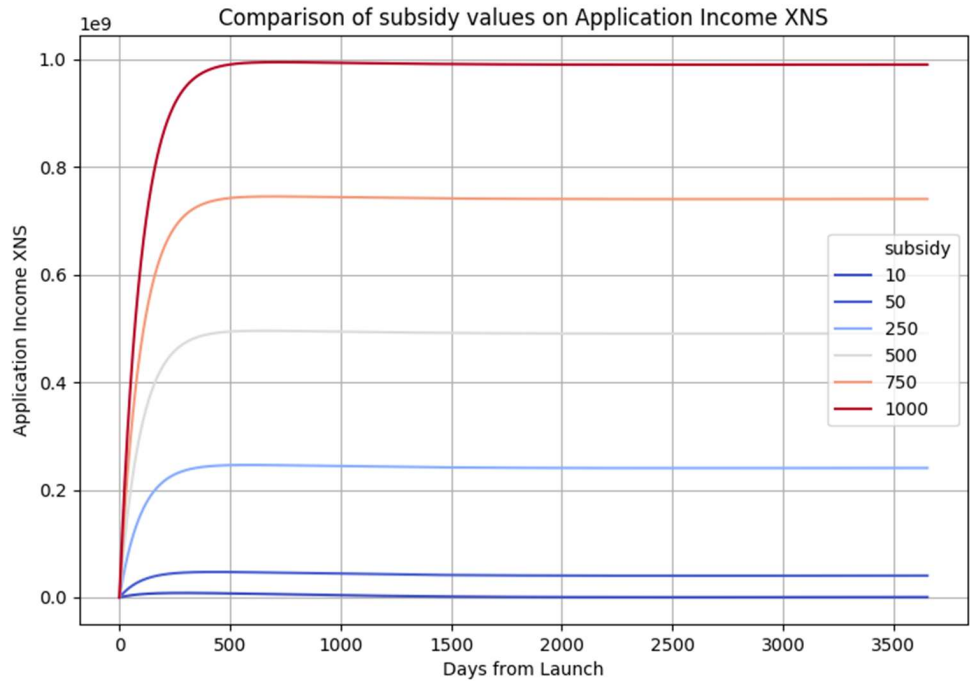

**Figure 11. The accumulated XNS tokens flowing to application developers – revised developer subsidy distribution strategy.**

Finally, in figure 12, we see that the changes to the way developer subsidies are distributed has a positive effect on the XNS price performance indicator. Initially, attention may be drawn to the fact that the benefits of allocating the largest number of XNS tokens to the pool (1000x10^6) have been significantly reduced - this would appear to be bad news at first. However, we also see that the smaller allocation of 200x10^6 XNS tokens now has a *positive* effect on the XNS price, whereas before, this allocation level resulting in a slight decrease in XNS price over 10 years. This is a great result for system designers. They are able to grow network value just by changing token allocation strategy, without additional resources!

The strategy which allocates 200x10^6 XNS tokens to the initial developer subsidy pool appears to outperform both higher and lower allocations (10, 50, 500, 750, and 1000). This suggests that there is an optimal number in the range of 50-250 which can be discovered through future work and experimentation using the validated simulation model.

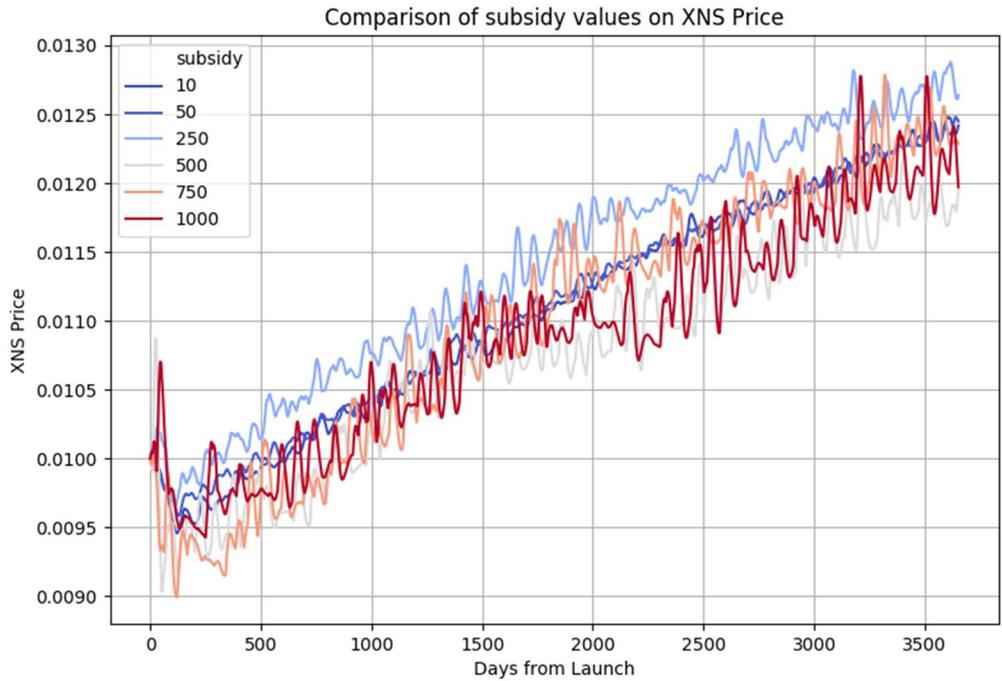

Figure 12. The predicted XNS price in dollars – revised developer subsidy distribution strategy

# 8  Conclusions

We elucidated a real-world example of how a Data Science process using Stochastic Dynamical Simulation and Modelling can be used in the context of designing cryptoeconomic systems. This work provides a rare view into the practical application of Data Science and Stochastic Simulation and Modelling to a Token Engineering design problem. We demonstrated how the described process has the potential to uncover previously unexpected factors in system level behaviours.

Specifically, after some analysis of initial simulation results it became clear that the first 2-3 years of adoption platform as an application platform were the most important. These results and subsequent analysis suggest that distributing the Application Developer Subsidy over a long time is inefficient, and it's much more effective to distribute it to Application Developers early on. These results further highlight the sensitivity of emergent outcomes to temporal dynamics as they are linked to path dependence properties in complex systems.

The suggested course of action is generally reasonable: the sooner an extremely compelling use case or "killer app" emerges, the greater chance the network will have generating "network effects" and accruing value. Based on results, a developer subsidy delivered early on has the best chance of accomplishing this. The refined Token Model was implemented in the "digital twin" or simulation model, and used to validate the idea that focusing application developer subsides early on in the network history was more efficient. Indeed, the new version of the Token model was shown to have several improved characteristics without requiring expenditure of additional resources.

Furthermore, it is observed that the Data Science and Simulation Modeling approach is valuable in that going through the process itself creates opportunities for the discovery of new knowledge and business understanding while developing the system from a high level specification to one precise enough to be executed as a computational model. Discovery of performance issues during design time can spare costly emergency interventions that would be necessary if issues instead became apparent in a production network. For this reason, network designers are increasingly adopting evidence-based design practices, such as the one described herein.